\documentclass[10pt]{iopartmod}
\usepackage{amssymb}
\usepackage{amsmath}
\usepackage{makeidx}
\usepackage[dvips]{graphicx}

\input{tcilatex}

\begin{document}
\title{Gigantic transmission band edge resonance in periodic stacks of anisotropic layers}
\author{Alex Figotin and Ilya Vitebskiy}
\affiliation{University of California at Irvine}

\begin{abstract}
We consider Fabry-Perot cavity resonance in periodic stacks of anisotropic
layers with misaligned in-plane anisotropy at the frequency close to a
photonic band edge. We show that in-plane dielectric anisotropy can result in
a dramatic increase in field intensity and group delay associated with the
transmission resonance. The field enhancement appears to be proportional to
forth degree of the number $N$ of layers in the stack. By contrast, in common
periodic stacks of isotropic layers, those effects are much weaker and
proportional to $N^{2}$. Thus, the anisotropy allows to drastically reduce the
size of the resonance cavity with similar performance. The key characteristic
of the periodic arrays with the gigantic transmission resonance is that the
dispersion curve $\omega\left(  k\right)  $ at the photonic band edge has the
degenerate form $\Delta\omega\sim\left(  \Delta k\right)  ^{4}$, rather than
the regular form $\Delta\omega\sim\left(  \Delta k\right)  ^{2}$. This can be
realized in specially arranged stacks of misaligned anisotropic layers. The
degenerate band edge cavity resonance with similar outstanding properties can
also be realized in a waveguide environment, as well as in a linear array of
coupled multimode resonators, provided that certain symmetry conditions are in place.

\end{abstract}
\maketitle

\section{Introduction}

The subject of this paper is the Fabry-Perot cavity resonance in periodic
layered structures. This phenomenon, known for many decades, is also referred
to as the \emph{transmission band edge resonance}, because it occurs in the
vicinity of photonic band edge frequencies in finite photonic crystals and is
accompanied by sharp transmission peaks. This basic effect can occur in a
finite periodic array of any two alternating materials with different
refractive indices. It can also be realized in a waveguide environment, or in
a finite array of coupled resonators. The only essential requirements are: (a)
low absorption, (b) the appropriate number $N$ of unit cells in the periodic
array, and (c) the presence of a frequency gap (a stop-band) in the frequency
spectrum of the periodic array. The latter is a universal property of almost
any lossless spatially periodic structure. This phenomenon has found numerous
and diverse practical applications in optics. Below we briefly outline some
basic features of the classical band edge resonance in periodic stacks of
isotropic layers. We only highlight those points, which are necessary for the
following comparative analysis of stacks involving anisotropic layers. More
detailed information on the relevant aspects of electrodynamics of layered
dielectric media can be found in an extensive literature on the subject (see,
for example, \cite{Yariv,Yeh,Chew,CR Meloni 03,Manda03,SL Scal1,SL Scal2,SL
Joann,CR Yariv04}, and references therein).

\begin{figure}[tbph]
\scalebox{0.8}{\includegraphics[viewport=0 0 500 250,clip]{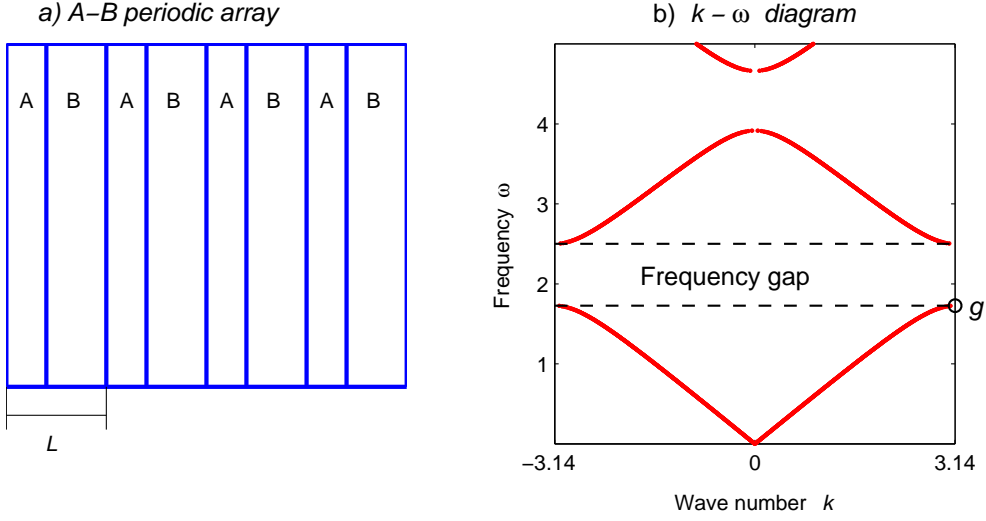}}
\caption{a) Periodic stack composed of two alternating layers $A$ and $B$,
each of which is made of isotropic transparent material. $L$ is the unit cell
of the periodic structure. The physical characteristics of the stack are
specified in the Appendix. b) The respective $k-\omega$ diagram. The
wavenumber $k$ and the frequency $\omega$ are expressed in units of $1/L$ and
$c/L$. The point $g$ at the Brillouin zone boundary designates the edge of the
lowest frequency band.}
\label{AB_DR}
\end{figure}

Consider a simplest periodic array of alternating dielectric layers $A$ and
$B$,\ as shown in Fig. \ref{AB_DR}(a). The layers are made of transparent
isotropic materials with different refractive indices $n_{A}$ and $n_{B}$.
Such periodic stacks are also referred to as a $1D$ photonic crystal. The
electromagnetic eigenmodes of the periodic structure in Fig. \ref{AB_DR}(a)
are Bloch waves with a typical wave number/frequency diagram shown in Fig.
\ref{AB_DR}(b).

Consider now a finite periodic stack composed of $N$ unit cells $L$ in Fig.
\ref{AB_DR}(a). Such a stack is commonly referred to as a Fabry-Perot cavity.
If the number $N$ of the double layers $L$ is significant, the stack
periodicity causes coherent interference of light scattered by the layers
interfaces. In practice, periodic stacks having as few as several periods $L$,
can display almost total reflectivity at the band gap frequencies, provided
that the refractive indices $n_{A}$ and $n_{B}$ of the adjacent layers differ significantly.%

\begin{figure}[tbph]
\scalebox{0.75}{\includegraphics[viewport=0 0 500 250,clip]{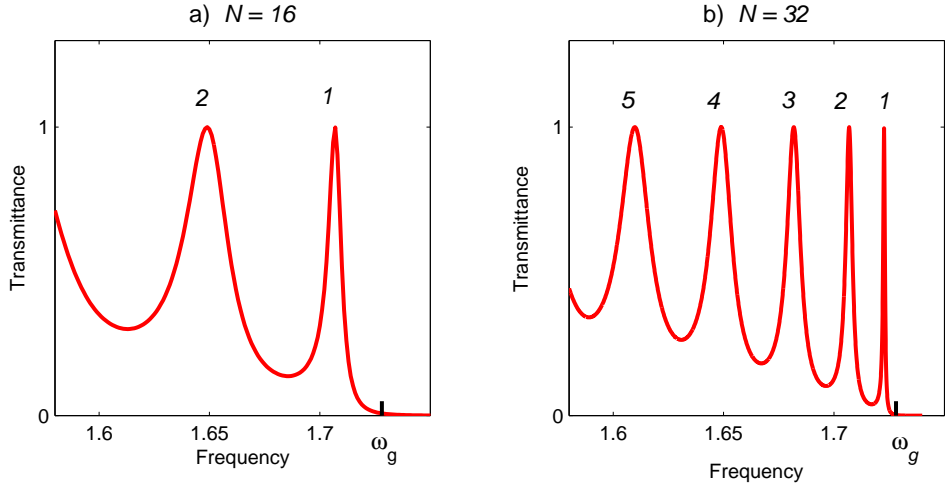}}
\caption{Typical transmission spectrum of finite periodic stacks composed of
different number $N$ of unit cells $L$ shown in Fig. 1(a). The sharp
transmission peaks below the band edge frequency $\omega_{g}$ are associated
with photonic band edge resonance.}
\label{tfN_iso}
\end{figure}

\begin{figure}[tbph]
\scalebox{0.75}{\includegraphics[viewport=0 0 500 250,clip]{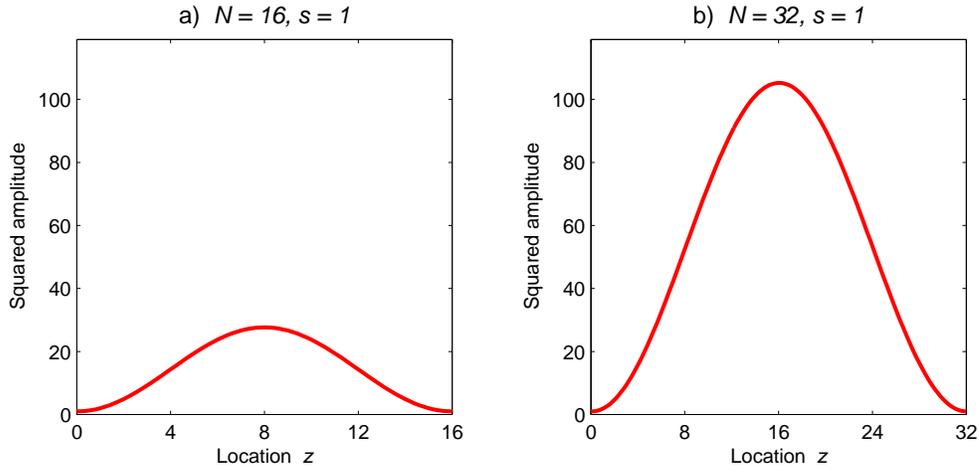}}
\caption{Smoothed intensity distribution (59) inside periodic stack at the
frequency $\omega_{1}\left(  N\right)  $ in (4) of the first transmission
resonance. The amplitude of the incident wave is unity. The distance $z$ from
the left boundary is expressed in units of $L$. At the stack boundaries at
$z=0$\ and $z=NL$, the field intensity is of the order of unity, while inside
the slab it is enhanced by factor $N^{2}$.}
\label{Amr1}
\end{figure}

\begin{figure}[tbph]
\scalebox{0.75}{\includegraphics[viewport=0 0 500 250,clip]{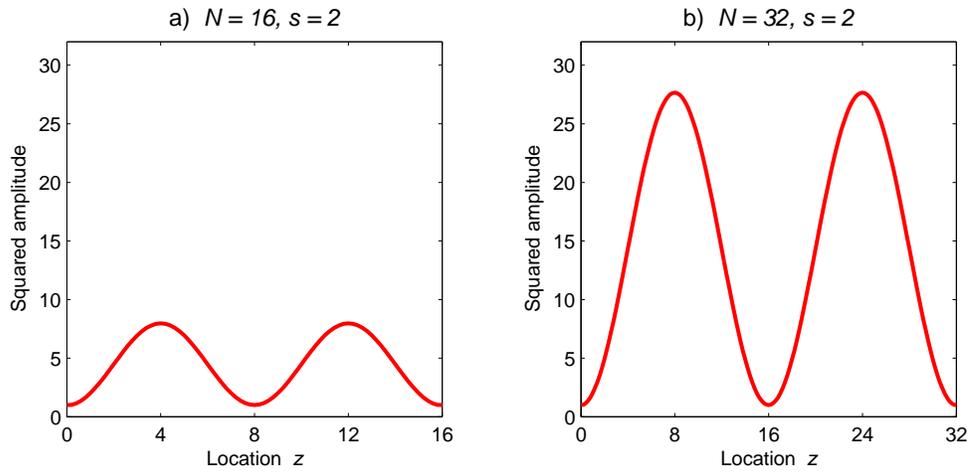}}
\caption{The same as in Fig. 3, but at the frequency $\omega_{2}\left(
N\right)  $ of the second transmission resonance in Figs. 2(a) and 2(b),
respectively.}
\label{Amr2}
\end{figure}

A typical frequency dependence of the finite stack transmittance is presented
in Fig. \ref{tfN_iso}. The frequency range shown includes the vicinity of the
photonic band edge (BE) $g$ in Fig. \ref{AB_DR}(b). The sharp transmission
peaks below the photonic band edge frequency $\omega_{g}$ correspond to
transmission band edge resonances, also known as Fabry-Perot cavity
resonances. At each resonance, the electromagnetic field inside the periodic
stack is close to a standing wave composed of a forward and a backward Bloch
eigenmodes with large and nearly equal amplitudes. The slab boundaries
coincide with the standing wave nodes, where the forward and backward Bloch
components interfere destructively, as illustrated in Figs. \ref{Amr1} and
\ref{Amr2}. The latter circumstance determines the wave numbers of the forward
and backward Bloch components at the resonance frequencies%
\begin{equation}
k_{s}\approx k_{g}\pm\frac{\pi}{NL}s,\ \ s=1,2,..., \label{k_s}%
\end{equation}
where $s$ is the order number of the resonant peak in Fig. \ref{tfN_iso}, and
$k_{g}$ is the wave number corresponding to the photonic band edge. In our
case, $k_{g}=\pi/L$. The resonance frequencies themselves can be expressed in
terms of the dispersion relation $\omega\left(  k\right)  $ of the respective
frequency band. Indeed, just below the photonic band edge $g$ in Fig.
\ref{AB_DR}(b), the dispersion relation $\omega\left(  k\right)  $ can be
approximated by the quadratic parabola
\begin{equation}
\omega\approx\omega_{g}+\frac{\omega_{g}^{\prime\prime}}{2}\left(
k-k_{g}\right)  ^{2},\text{ \ where }\omega_{g}^{\prime\prime}=\left(
\frac{\partial^{2}\omega}{\partial k^{2}}\right)  _{k=k_{g}}<0.
\label{w = k^2}%
\end{equation}
This relation together with (\ref{k_s}) yield the frequencies of the resonant
transmission peaks in Fig. \ref{tfN_iso}%
\begin{equation}
\omega_{s}\left(  N\right)  \approx\omega_{g}+\frac{\omega_{g}^{\prime\prime}%
}{2}\left(  \frac{\pi}{NL}s\right)  ^{2},\ \ s=1,2,... \label{w_s}%
\end{equation}
where $\omega_{g}=\omega\left(  k_{g}\right)  $ is the band edge frequency.
For example, the transmission peak $1$ closest to the photonic band edge is
located at%
\begin{equation}
\omega_{1}\left(  N\right)  \approx\omega_{g}+\frac{\omega_{g}^{\prime\prime}%
}{2}\left(  \frac{\pi}{NL}\right)  ^{2}. \label{w_1}%
\end{equation}
The dependence (\ref{w_s}) is illustrated in Fig. \ref{tfN_iso}.%

Electromagnetic field distribution inside the stack at the frequency of band
edge resonance is shown in Figs. \ref{Amr1} and \ref{Amr2} for the first two
transmission resonances, respectively. For a given amplitude $\Psi_{I}$ of the
incident wave, the maximal field intensity $\left\vert \Psi\left(  z\right)
\right\vert ^{2}$ inside the slab depends on the number $N$ of the double
layers in the stack and on the order number $s$ of the resonance peak in Fig.
\ref{tfN_iso}%
\begin{equation}
\max\left\vert \Psi\left(  z\right)  \right\vert ^{2}\propto\left\vert
\Psi_{I}\right\vert ^{2}\left(  \frac{N}{s}\right)  ^{2}. \label{W_g}%
\end{equation}
The maximal field intensity (\ref{W_g}) is proportional to squared thickness
of the slab and, for a large $N$, is greatly enhanced compared to that of the
incident light. By contrast, at the slab boundaries at $z=0$ and $z=D=NL$, the
field amplitude $\Psi\left(  z\right)  $ always remains comparable to
$\Psi_{I}$ to satisfy the electromagnetic boundary conditions (\ref{BC2}). At
frequencies outside the resonance transparency peaks, the field amplitude
inside the stack drops sharply.

The exact definition of the physical values plotted in Figs. \ref{Amr1} and
\ref{Amr2}, as well as the numerical parameters of the periodic stacks used to
generates these plots, are given in the Appendix.%

The expressions (\ref{k_s}) through ((\ref{W_g}) are valid if $N\gg1$ and only
apply to the transmission resonances close enough to the photonic band edge.
In further consideration, we will focus on the most powerful first resonance
$s=1$, closest to the photonic band edge.

\begin{figure}[tbph]
\scalebox{0.8}{\includegraphics[viewport=0 0 350 180,clip]{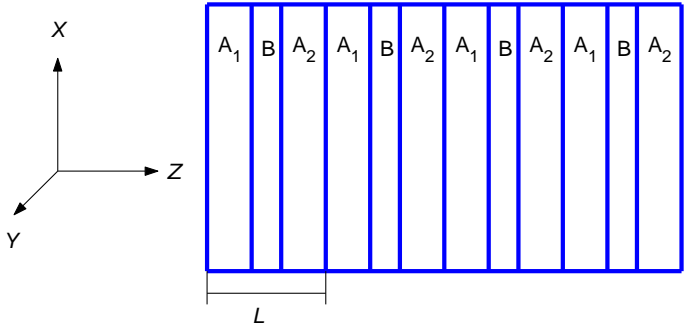}}
\caption{Periodic stack capable of supporting $k-\omega$ diagram with
degenerate band edge. A unit cell $L$ includes three layers: two birefringent
layers $A_{1}$ and $A_{2}$ with different orientations $\varphi_{1}$ and
$\varphi_{2}$ of the respective anisotropy axes in the $X-Y$ plane, and one
isotropic $B$ layer. The misalignment angle $\varphi=\varphi_{1}-\varphi_{2}$
between adjacent $A$ layers must be different from $0$ and $\pi/2$.}
\label{ABA}
\end{figure}

\begin{figure}[tbph]
\scalebox{0.75}{\includegraphics[viewport=0 0 500 400,clip]{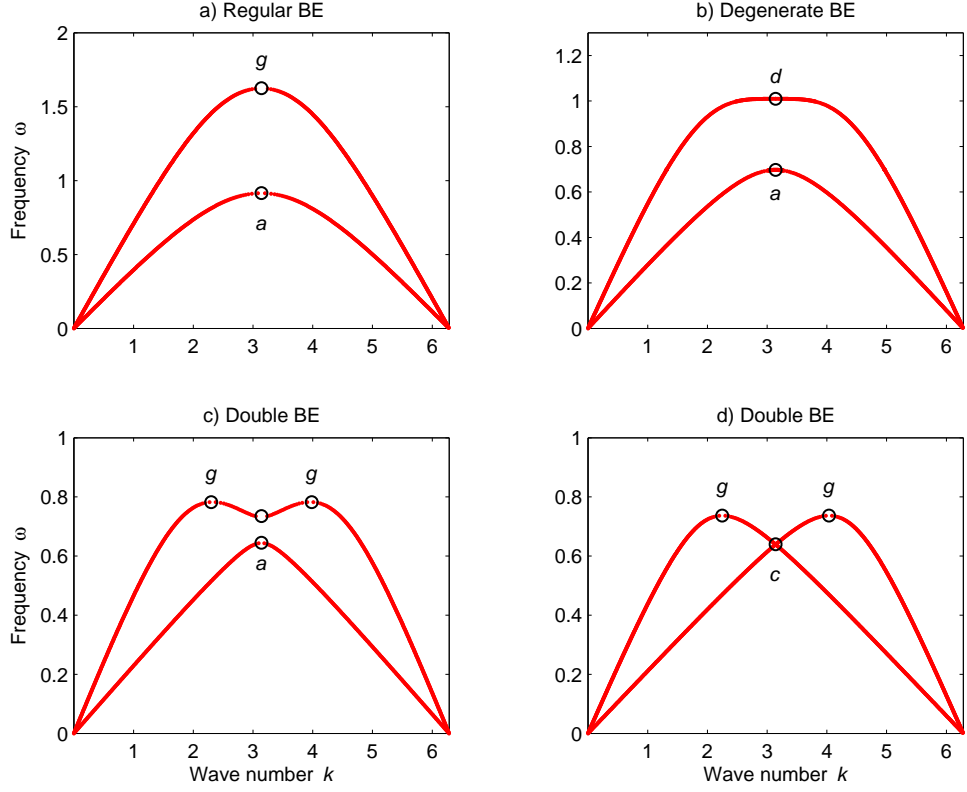}}
\caption{The first band of the $k-\omega$ diagram of the periodic stack in
Fig. 5 for four different values of the $B$ - layer thickness, $D_{B}$. (a)
$D_{B}=0.75891\times L$. (b) $D_{B}=0.45891\times L$; in this case the upper
dispersion curve develops degenerate band edge $d$. (c) $D_{B}=0.35891\times L$.
(d) $D_{B}=0$; in this case the two intersecting dispersion curves correspond to
the Bloch waves with different symmetries, and the respective modes are
decoupled. The physical parameters of the periodic structure are specified in
the Appendix.}
\label{DRd_4}
\end{figure}

Above, we outlined some basic features of the transmission band edge resonance
in finite stacks of isotropic layers. The question we would like to address in
this paper is whether the presence of anisotropic layers in a finite periodic
stack can qualitatively change the nature of the Fabry-Perot cavity resonance.
We will show that, indeed, in periodic stacks involving anisotropic layers,
the transmission resonance can be significantly stronger, compared to what is
achievable with common periodic stacks of isotropic layers. For instance, in
the periodic stack shown in Fig. \ref{ABA}, the field intensity associated
with the transmission band edge resonances can be proportional to $N^{4}$,
rather than $N^{2}$. The latter implies that a stack of $N$ anisotropic layers
can perform as well as a common stack of $N^{2}$ isotropic layers. And this is
a huge difference! The physical reason for this is that periodic stacks of
anisotropic layers can support the kind of $k-\omega$ diagrams that are
impossible in stacks of isotropic layers. Specifically, a dispersion curve
$\omega\left(  k\right)  $ of the stack in Fig. \ref{ABA} can develop a
degenerate band edge, as shown in Fig. \ref{DRd_4}(b). Just below the
degenerate band edge $d$, the dispersion curve can be approximated as%
\begin{equation}
\omega\approx\omega_{d}+\frac{\omega_{d}^{\prime\prime\prime\prime}}%
{24}\left(  k-k_{d}\right)  ^{4},\ \text{where }\ \omega_{d}^{\prime
\prime\prime\prime}=\left(  \frac{\partial^{4}\omega}{\partial k^{4}}\right)
_{k=k_{d}}<0, \label{w = k^4}%
\end{equation}
which implies a huge density of modes. And this is what makes all the
difference compared to the case (\ref{w = k^2}) of a regular band edge.%

For convenience, the domain of definition of the Bloch wave number $k$ in Fig.
\ref{DRd_4} is chosen between $0$ and $2\pi/L$. Since the Bloch wave number is
defined up to a multiple of $2\pi/L$, the representation in Fig. \ref{DRd_4}
is equivalent to that in Fig. \ref{AB_DR}(b). Note that the points $a$, $g$,
and $d$ in Fig. \ref{DRd_4}(a) and \ref{DRd_4}(b) lie at the Brillouin zone
boundary at $k=\pi/L$.

\begin{figure}[tbph]
\scalebox{0.75}{\includegraphics[viewport=0 0 500 180,clip]{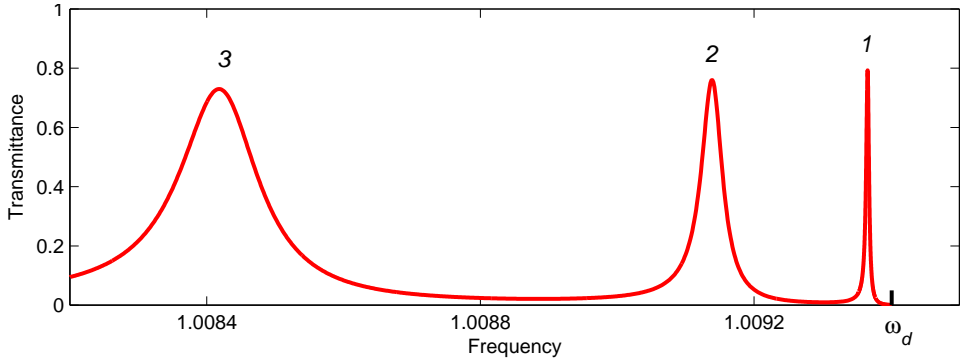}}
\caption{The transmission spectrum of periodic stack with the $k-\omega$
diagram in Fig. 6(b). The stack is composed of $N=32$ three-layered unit cells
$L$ in Fig. 5. The sharp transmission peaks below the degenerate band edge
frequency $\omega_{d}$ are associated with Fabry-Perot cavity resonances.}
\label{tf32yf}
\end{figure}

\begin{figure}[tbph]
\scalebox{0.75}{\includegraphics[viewport=0 0 500 250,clip]{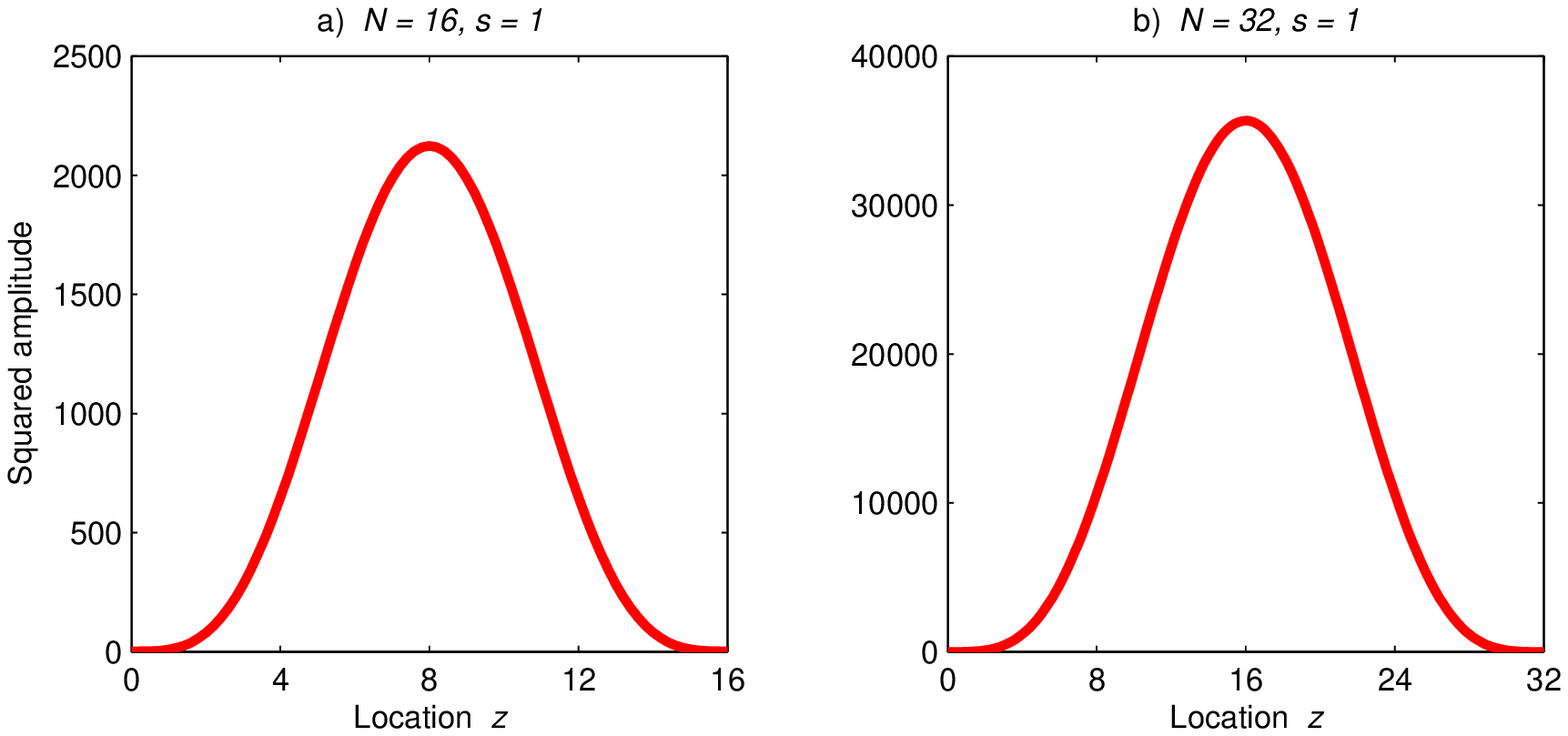}}
\caption{Smoothed intensity distribution inside periodic stack in Fig. 5 at
the frequency of the first Fabry-Perot resonance, closest to degenerate band
edge $d$. The amplitude of the incident wave is unity. At the stack boundaries
at $z=0$\ and $z=NL$, the field intensity is of the order of unity, while
inside the slab electromagnetic energy density is enhanced by factor $N^{4}$.}
\label{Amdny}
\end{figure}

In Fig. \ref{tf32yf} we present the transmission dispersion of the finite
periodic stacks in Fig. \ref{ABA} composed of 32 unit cells $L$ and having the
$k-\omega$ diagram in Fig. \ref{DRd_4}(b). The frequency range shown includes
the degenerate band edge (DBE) at $\omega=\omega_{d}$. The field intensity
distribution at the frequency of the first transmission resonance 1 is shown
in Fig. \ref{Amdny}(b). One can see that for a given $N$, the resonant field
intensity in Fig. \ref{Amdny} is significantly larger compared to that of the
vicinity of a regular band edge, shown in Fig. \ref{Amr1}. Specifically, in
the case of degenerate band edge%
\begin{equation}
\max\left\vert \Psi\left(  z\right)  \right\vert ^{2}\propto\left\vert
\Psi_{I}\right\vert ^{2}\left(  \frac{N}{s}\right)  ^{4}, \label{W_d}%
\end{equation}
compared to the estimation (\ref{W_g}) related to a regular band edge. The
transmission bandwidth in the case of DBE appears to be much smaller.%

In practice, the field amplitude associated with the transmission resonance is
limited not only by the number of layers in the stack, but also by such
factors as absorption, nonlinearity, imperfections of the periodic array,
stack dimensions in the $X-Y$ plane, incident radiation bandwidth, etc. All
else being equal, the stack with degenerate band edge can have much fewer
layers and, therefore, can be much thinner compared to a regular stack of
isotropic layers with similar performance. Much smaller dimensions can be very
attractive for a variety of practical applications. Similar effect associated
with degenerate band edge (\ref{w = k^4}) of the $k-\omega$ diagram can also
be achieved in the waveguide environment, as well as in finite periodic arrays
of coupled multiple-mode resonators. On the down side, the realization of
$k-\omega$ diagram having a dispersion curve with DBE, requires more
sophisticated periodic arrays, such as the one shown in Fig. \ref{ABA}.

The rest of the paper is organized as follows.

In Section 2 we briefly outline the electrodynamics of periodic stratified
media and examine the relation between anisotropy of the layers and the
$k-\omega$ diagram of the stack. We show that at the frequency of degenerate
band edge $d$, the $4\times4$ transfer matrix of a unit cell $L$ cannot be
diagonalized or even reduced to the block-diagonal form. Based on this, we
establish necessary symmetry conditions for a periodic stack to develop a
degenerate band edge (\ref{w = k^4}) and to display the peculiar resonance
properties associated with it. We prove that such periodic stacks must have at
least two misaligned anisotropic layers in a unit cell, as shown in the
example in Fig. \ref{ABA}.

In Section 3, we consider the scattering problem for a finite periodic stack.
We analyze the eigenmode composition of electromagnetic field at the frequency
of transmission resonance near degenerate band edge. We show that in contrast
to the case (\ref{W_g}) of a regular band edge, in the degenerate band edge
case (\ref{W_d}) the resonance field inside the stack does not reduce to a
superposition of forward and backward propagating eigenmodes. Instead, the
contribution of evanescent eigenmodes becomes equally important and leads to
much stronger dependence (\ref{W_d}) of the resonance field intensity on the
number of layers in the stack.

The physical and geometrical parameters of stacks used for numerical
simulations are specified in the Appendix.

\section{Electrodynamics of periodic stacks of anisotropic layers}

This section starts with a brief description of some basic electrodynamic
properties of periodic layered media composed of lossless anisotropic layers.
Then we turn to the particular case of periodic stacks with degenerate band
edge. The scattering problem for periodic finite stacks, including the
Fabry-Perot cavity resonance in the vicinity of degenerate photonic band edge
will be considered in the next section.

\subsection{Transverse electromagnetic waves in stratified media}

Our consideration is based on time-harmonic Maxwell equations in heterogeneous
nonconducting media%
\begin{equation}
\nabla\times\mathbf{\vec{E}}\left(  \vec{r}\right)  =i\frac{\omega}%
{c}\mathbf{B}\left(  \vec{r}\right)  ,\;\nabla\times\mathbf{\vec{H}}\left(
\vec{r}\right)  =-i\frac{\omega}{c}\mathbf{D}\left(  \vec{r}\right)  ,
\label{THME}%
\end{equation}
where electric and magnetic fields and inductions are related by linear
constitutive equations%
\begin{equation}
\mathbf{\vec{D}}\left(  \vec{r}\right)  =\hat{\varepsilon}\left(  \vec
{r}\right)  \mathbf{\vec{E}}\left(  \vec{r}\right)  ,\ \mathbf{\vec{B}}\left(
\vec{r}\right)  =\hat{\mu}\left(  \vec{r}\right)  \mathbf{\vec{H}}\left(
\vec{r}\right)  . \label{MR}%
\end{equation}

In further consideration we assume that:

\begin{enumerate}
\item The direction of plane wave propagation coincide with the normal $z$ to
the layers.

\item The second rank tensors $\hat{\varepsilon}\left(  \vec{r}\right)  $ and
$\hat{\mu}\left(  \vec{r}\right)  $ are dependent on a single Cartesian
coordinate $z$, normal to the layers.

\item The $z$ direction is a two-fold symmetry axis of the stack, implying
that the anisotropy axes of individual layers are either parallel, or
perpendicular the $z$ direction. The latter defines the case of in-plane anisotropy.
\end{enumerate}

Under the above restrictions, the normal field components $E_{z}$ and $H_{z}$
of electromagnetic wave are zeroes, and the system (\ref{THME}) of six
time-harmonic Maxwell equations reduces to the following system of four
ordinary linear differential equations for the transverse field components%

\begin{equation}
\frac{\partial}{\partial z}\Psi\left(  z\right)  =i\frac{\omega}{c}M\left(
z\right)  \Psi\left(  z\right)  ,\text{\ where }\Psi\left(  z\right)  =\left[
\begin{array}
[c]{c}%
E_{x}\left(  z\right) \\
E_{y}\left(  z\right) \\
H_{x}\left(  z\right) \\
H_{y}\left(  z\right)
\end{array}
\right]  . \label{ME4}%
\end{equation}
The $4\times4$ matrix $M\left(  z\right)  $ in (\ref{ME4}) is referred to as
the (reduced) Maxwell operator.

In a lossless nonmagnetic medium with in-plane anisotropy, the electric
permittivity and magnetic permeability tensors have the form%
\begin{equation}
\hat{\varepsilon}=\left[
\begin{array}
[c]{ccc}%
\varepsilon_{xx} & \varepsilon_{xy} & 0\\
\varepsilon_{xy} & \varepsilon_{yy} & 0\\
0 & 0 & \varepsilon_{zz}%
\end{array}
\right]  ,\ \hat{\mu}=\hat{1}. \label{eps trnsv}%
\end{equation}
This yields the following explicit expression for the Maxwell operator
$M\left(  z\right)  $ in (\ref{ME4})%
\begin{equation}
M\left(  z\right)  =\left[
\begin{array}
[c]{cccc}%
0 & 0 & 0 & 1\\
0 & 0 & -1 & 0\\
-\varepsilon_{xy} & -\varepsilon_{yy} & 0 & 0\\
\varepsilon_{xx} & \varepsilon_{xy} & 0 & 0
\end{array}
\right]  , \label{M trnsv}%
\end{equation}
where the components of the permittivity tensor may vary from layer to layer.

\subsubsection{The transfer matrix formalism}

The Cauchy problem%
\begin{equation}
\frac{\partial}{\partial z}\Psi\left(  z\right)  =i\frac{\omega}{c}M\left(
z\right)  \Psi\left(  z\right)  ,\;\Psi\left(  z_{0}\right)  =\Psi_{0}
\label{CauPsi}%
\end{equation}
for the reduced Maxwell equation (\ref{ME4}) has a unique solution%
\begin{equation}
\Psi\left(  z\right)  =T\left(  z,z_{0}\right)  \Psi\left(  z_{0}\right)  ,
\label{T(zz0)}%
\end{equation}
where the $4\times4$ matrix $T\left(  z,z_{0}\right)  $ is referred to as the
\emph{transfer matrix}. The transfer matrix (\ref{T(zz0)}) uniquely relates
the values of time-harmonic electromagnetic field $\Psi$ at any two points $z$
and $z_{0}$ of the stratified medium. From the definition (\ref{T(zz0)}), it
follows that%
\begin{equation}
T\left(  z,z_{0}\right)  =T\left(  z,z^{\prime}\right)  T\left(  z^{\prime
},z_{0}\right)  ,\;T\left(  z,z_{0}\right)  =T^{-1}\left(  z_{0},z\right)
,\;T\left(  z,z\right)  =I. \label{Tzz0}%
\end{equation}

The transfer matrix of a stack of layers is defined as%
\[
T_{S}=T\left(  D,0\right)  ,
\]
where $z=0$ and $z=D$ are the stack boundaries. The greatest advantage of the
transfer matrix formalism stems from the fact that the transfer matrix of an
arbitrary stack is a sequential product of the transfer matrices $T_{m}$ of
the constitutive layers%
\begin{equation}
T_{S}=\prod_{m}T_{m}. \label{TS}%
\end{equation}
If the individual layers $m$ are homogeneous, the corresponding single-layer
transfer matrices $T_{m}$ can be explicitly expressed in terms of the
respective Maxwell operators $M_{m}$%
\begin{equation}
T_{m}=\exp\left(  iD_{m}M_{m}\right)  , \label{Tm}%
\end{equation}
where $D_{m}$ is the thickness of the $m$-th layer. The explicit expression
for the Maxwell operator $M_{m}$ of a uniform dielectric layer with in-plane
anisotropy is given by Eq. (\ref{M trnsv}). Thus, Eq. (\ref{TS}) together with
(\ref{Tm}) and (\ref{M trnsv}) provide an explicit analytical expression for
the transfer matrix $T_{S}$ of a stack of dielectric layers with in-plane anisotropy.

The $4\times4$ transfer matrix of an arbitrary lossless stratified medium
displays the fundamental property of $J$-unitarity \cite{PRE03}%
\begin{equation}
T^{\dag}=JT^{-1}J,\text{\ \ where \ }J=\left[
\begin{array}
[c]{cccc}%
0 & 0 & 0 & 1\\
0 & 0 & -1 & 0\\
0 & -1 & 0 & 0\\
1 & 0 & 0 & 0
\end{array}
\right]  , \label{JU}%
\end{equation}
which implies, in particular, that%
\begin{equation}
\left\vert \det T\right\vert =1. \label{detT=1}%
\end{equation}

Different versions of the $4\times4$ transfer matrix formalism have been used
in electrodynamics of stratified media composed of birefringent and/or
gyrotropic layers for decades (see, for example,
\cite{Tmatrix,Abdul99,Abdul00} and references therein). In this paper we use
exactly the same notations and terminology as in our previous publications
\cite{PRE01,PRB03,PRE03,PRE05} on electrodynamics of stratified media.

\subsection{Eigenmodes in periodic layered media}

In a periodic layered medium, all material tensors are periodic functions of
$z$, and so is the $4\times4$ matrix $M(z)$ in (\ref{ME4}). Usually, the
solutions $\Psi_{k}\left(  z\right)  $ of the reduced Maxwell equation
(\ref{ME4}) with the periodic $M(z)$ can be chosen in the Bloch form%
\begin{equation}
\Psi_{k}\left(  z+L\right)  =e^{ikL}\Psi_{k}\left(  z\right)  , \label{Bloch}%
\end{equation}
where the Bloch wave number $k$ is defined up to a multiple of $2\pi/L$. The
definition (\ref{T(zz0)}) of the transfer matrix together with Eq.
(\ref{Bloch}) yield%
\begin{equation}
T\left(  z+L,z\right)  \Psi_{k}\left(  z\right)  =e^{ikL}\Psi_{k}\left(
z\right)  . \label{Psi=TPsi}%
\end{equation}
Introducing the transfer matrix of a unit cell $L$%
\begin{equation}
T_{L}=T\left(  L,0\right)  , \label{TL}%
\end{equation}
we have from Eq. (\ref{Psi=TPsi})%
\begin{equation}
T_{L}\Phi_{k}=e^{ikL}\Phi_{k},\text{ \ \ where }\;\Phi_{k}=\Psi_{k}\left(
0\right)  . \label{T(L)=e(ikL)}%
\end{equation}
Thus, the eigenvectors of the transfer matrix $T_{L}$ of the unit cell are
uniquely related to the Bloch solutions $\Psi_{k}\left(  z\right)  $ of the
reduced Maxwell equation (\ref{ME4})%
\begin{equation}
\Phi_{i}=\Psi_{i}\left(  0\right)  ,\;i=1,2,3,4. \label{Psi 1234}%
\end{equation}
The respective four eigenvalues%
\begin{equation}
X_{i}=e^{ik_{i}L},\;i=1,2,3,4 \label{X(k)}%
\end{equation}
of $T_{L}$ are the roots of the characteristic polynomial $F_{4}\left(
X\right)  $ of the forth degree%
\begin{equation}
F_{4}\left(  X\right)  =\det\left(  T_{L}-XI\right)  =0. \label{Char X}%
\end{equation}

Unit cell of a periodic stack can be chosen differently. For example, the
choice $A_{1}-B-A_{2}$ specified in Fig. \ref{ABA} is as good as $A_{2}%
-A_{1}-B$. Different choice of a unit cell corresponds to its shift in the $z$
direction and results in the following transformation of the respective
transfer matrix $T_{L}$%
\begin{equation}
T_{L}^{\prime}=T\left(  0,Z\right)  T_{L}T\left(  Z,0\right)  =\left[
T\left(  Z,0\right)  \right]  ^{-1}T_{L}T\left(  Z,0\right)  , \label{Shift}%
\end{equation}
where $Z$ is the amount of the shift. The modified transfer matrix is similar
to the original one and has the same set of eigenvalues.

For any given $\omega$, the characteristic equation (\ref{Char X}) defines a
set of four eigenvalues (\ref{X(k)}). Real $k$ (or, equivalently, $\left\vert
X\right\vert =1$) correspond to propagating Bloch modes, while complex $k$
(or, equivalently, $\left\vert X\right\vert \neq1$) correspond to evanescent modes.

The $J$-unitarity (\ref{JU}) of $T_{L}$ imposes the following restriction on
its eigenvalues (\ref{X(k)})%
\begin{equation}
\left\{  X_{i}^{-1}\right\}  \equiv\left\{  X_{i}^{\ast}\right\}  ,\;i=1,2,3,4
\label{X*=1/X}%
\end{equation}
or, equivalently%
\begin{equation}
\{k_{i}\}\equiv\{k_{i}^{\ast}\},\;i=1,2,3,4, \label{k=k*}%
\end{equation}
for any given $\omega$. In view of the relations (\ref{k=k*}) or
(\ref{X*=1/X}), one can distinguish the following three different situations.

\begin{enumerate}
\item[A.] All four wave numbers are real%
\begin{equation}
k_{1}\equiv k_{1}^{\ast},\;k_{2}\equiv k_{2}^{\ast},\;k_{3}\equiv k_{3}^{\ast
},\;k_{4}\equiv k_{4}^{\ast}. \label{4ex}%
\end{equation}
In the example in Fig. \ref{DRd_4}, this case relates to the frequency range%
\begin{equation}
0<\omega<\omega_{a}. \label{w<wa}%
\end{equation}
In this case, all four Bloch eigenmodes are propagating.

\item[B.] Two wave numbers are real and the other two are complex%
\begin{equation}
k_{1}=k_{1}^{\ast},\;k_{2}=k_{2}^{\ast},\;k_{4}=k_{3}^{\ast},\;\text{where
}k_{3}\neq k_{3}^{\ast},\;k_{4}\neq k_{4}^{\ast}. \label{2ex2ev}%
\end{equation}
This case relates to the frequency range
\begin{equation}
\omega_{a}<\omega<\omega_{g}, \label{wa<w<wg}%
\end{equation}
in Fig. \ref{DRd_4}(a), or the frequency range%
\begin{equation}
\omega_{a}<\omega<\omega_{d}, \label{wa<w<wd}%
\end{equation}
in Fig. \ref{DRd_4}(b). In both cases (\ref{wa<w<wg}) and (\ref{wa<w<wd}), two
of the four Bloch eigenmodes are propagating and the remaining two are
evanescent with complex conjugated wave numbers.

\item[C.] All four wave numbers are complex%
\begin{equation}
k_{2}=k_{1}^{\ast},\;k_{4}=k_{3}^{\ast},\;\text{where }k_{1}\neq k_{1}^{\ast
},\;k_{2}\neq k_{2}^{\ast},\;k_{3}\neq k_{3}^{\ast},\;k_{4}\neq k_{4}^{\ast}.
\label{4ev}%
\end{equation}
This situation relates to a frequency gap, where all four Bloch eigenmodes are
evanescent. In the example in Fig. \ref{DRd_4}, this case corresponds to%
\begin{equation}
\omega_{g}<\omega,\text{ \ or \ \ }\omega_{d}<\omega. \label{wg<w}%
\end{equation}

\end{enumerate}

Notice that the case (B) of two propagating and two evanescent modes can only
occur in periodic stacks of anisotropic layers. If all the layers in a unit
cell are isotropic, the four Bloch eigenmodes are either all propagating or
all evanescent, as is the case in Fig. \ref{AB_DR}. For a given frequency
$\omega$, the four Bloch eigenmodes correspond to two different polarizations
and two opposite directions of propagation.

\subsubsection{Non-Bloch solutions at stationary points of the $k-\omega$
diagram}

So far, we have considered only the cases where all four solutions for the
reduced Maxwell equation (\ref{ME4}) can be chosen in the Bloch form
(\ref{Bloch}). All such cases fall into one of the following three categories:
(A) all four eigenmodes are propagating, (B) two modes are propagating and the
other two are evanescent, (C) all four eigenmodes are evanescent. This
classification of the eigenmodes does not apply at the frequencies of
stationary points on the $k-\omega$ diagram, where the group velocity $u$ of
some of the propagating modes vanishes%
\begin{equation}
u=d\omega/dk=0. \label{SP}%
\end{equation}
At a stationary point of a dispersion curve, not all four solutions of the
Maxwell equation (\ref{ME4}) are Bloch waves, as defined in (\ref{Bloch}).
Instead, some of the solutions can be algebraically diverging non-Bloch
eigenmodes. Such eigenmodes can be essential for understanding the resonance
effects in finite and semi-infinite periodic arrays.

Let us consider the situation at stationary points (\ref{SP}) on the
$k-\omega$ diagram in terms of the matrix $T_{L}$. Although at any given
frequency $\omega$, the reduced Maxwell equation (\ref{ME4}) has exactly four
linearly independent solutions, it does not imply that the respective transfer
matrix $T_{L}$ in (\ref{T(L)=e(ikL)}) must have four linearly independent
eigenvectors (\ref{T(L)=e(ikL)}). Indeed, although the matrix $T_{L}$ is
invertible, it is neither Hermitian, nor unitary and, therefore, may not be
diagonalizable. Specifically, if the frequency approaches one of the
stationary points (\ref{SP}), some of the eigenvectors $\Phi_{k}$ in
(\ref{T(L)=e(ikL)}) become nearly parallel to each other. Eventually, as
$\omega$ reaches the the stationary point value, the number of linearly
independent eigenvectors $\Phi_{k}$ becomes lesser than four, and the relation
(\ref{Psi 1234}) does not apply at that point.

Examples of different stationary points (\ref{SP}) are shown in Fig.
\ref{DRd_4}. Using these examples, let us take a closer look at these special frequencies.

At the frequencies $\omega_{g}$ of the photonic band edge $g$ in Fig.
\ref{DRd_4}(a), the four eigenmodes include:

\begin{itemize}
\item[-] one propagating mode with $k=\pi/L$ and zero group velocity,

\item[-] one non-Bloch eigenmode linearly diverging with $z$,

\item[-] a pair of evanescent modes with equal and opposite imaginary wave numbers.
\end{itemize}

At the frequencies $\omega_{a}$ corresponding to the point $a$ in Fig.
\ref{DRd_4}, the four solutions of Eq. (\ref{ME4}) include:

\begin{itemize}
\item[-] the propagating mode with $k=\pi/L$ and zero group velocity,

\item[-] one non-Bloch eigenmode linearly diverging with $z$,

\item[-] a pair of propagating modes having equal and opposite group
velocities and belonging to the dispersion curve other than the one containing
the point $a$.
\end{itemize}

Of special interest here is the frequency $\omega_{d}$ of degenerate photonic
band edge $d$ in Fig. \ref{DRd_4}(b). In this case, the four solutions of Eq.
(\ref{ME4}) include \cite{WRM}:

\begin{itemize}
\item[-] the propagating mode with $k=\pi/L$ and zero group velocity,

\item[-] three non-Bloch eigenmodes diverging as $z$, $z^{2}$, and $z^{3}$, respectively.
\end{itemize}

At any particular frequency, the existence of non-Bloch eigenmodes can be
directly linked to the canonical Jordan form of the respective transfer matrix
$T_{L}$. Indeed, the $4\times4$ matrix $T_{L}$, being invertible, can have one
of the following five different canonical forms%
\begin{align}
\tilde{T}_{1}  &  =\left[
\begin{array}
[c]{cccc}%
X_{1} & 0 & 0 & 0\\
0 & X_{2} & 0 & 0\\
0 & 0 & X_{3} & 0\\
0 & 0 & 0 & X_{4}%
\end{array}
\right]  ,\nonumber\\
\tilde{T}_{21}  &  =\left[
\begin{array}
[c]{cccc}%
X_{1} & 1 & 0 & 0\\
0 & X_{1} & 0 & 0\\
0 & 0 & X_{3} & 0\\
0 & 0 & 0 & X_{4}%
\end{array}
\right]  ,\ \tilde{T}_{22}=\left[
\begin{array}
[c]{cccc}%
X_{1} & 1 & 0 & 0\\
0 & X_{1} & 0 & 0\\
0 & 0 & X_{2} & 1\\
0 & 0 & 0 & X_{2}%
\end{array}
\right]  ,\label{Jordan}\\
\tilde{T}_{3}  &  =\left[
\begin{array}
[c]{cccc}%
X_{1} & 1 & 0 & 0\\
0 & X_{1} & 1 & 0\\
0 & 0 & X_{1} & 0\\
0 & 0 & 0 & X_{2}%
\end{array}
\right]  ,\ \tilde{T}_{4}=\left[
\begin{array}
[c]{cccc}%
X & 1 & 0 & 0\\
0 & X & 1 & 0\\
0 & 0 & X & 1\\
0 & 0 & 0 & X
\end{array}
\right]  ,\nonumber
\end{align}
of which all but $\tilde{T}_{1}$ have non-trivial Jordan blocks and,
therefore, are not diagonalizable. Each $m\times m$ Jordan block is associated
with a single eigenvector of the respective $T$ - matrix. Therefore, the total
number of eigenvectors of different $4\times4$ matrix in (\ref{Jordan}) is
lesser than the number four of the solutions for the reduced Maxwell equation
(\ref{ME4}). The only exception is the diagonalizable case $\tilde{T}_{1}$,
where the four $T_{L}$ eigenvectors correspond to four Bloch eigenmodes, as
prescribed by (\ref{Psi 1234}). Generally, each nontrivial $m\times m$ Jordan
block of the $T$ - matrix is associated with the $m$ eigenmodes of Eq.
(\ref{ME4}), of which one is a propagating Bloch eigenmode with zero group
velocity, and the other $m-1$ are non-Bloch eigenmodes algebraically diverging
with $z$. The details can be found in any course on linear algebra, for
example, \cite{LODE,Linalg,LanTi,Wilk}. Let us consider each case separately.

At general frequencies, different from those of stationary points (\ref{SP}),
the $4\times4$ matrix $T_{L}$ is always diagonalizable. Its canonical
(diagonalized) form is trivial and coincides with $\tilde{T}_{1}$ in
(\ref{Jordan}). The four eigenvectors are defined in (\ref{Psi 1234}). All the
possibilities here reduce to one of the three cases (A), (B), or (C) described
earlier in this Section. None of them involves non-Bloch solutions.

At the frequency $\omega_{g}$ of a regular photonic band edge in Fig.
\ref{DRd_4}(a), the canonical Jordan form of the respective transfer matrix is
$\tilde{T}_{21}$ in (\ref{Jordan}), where%
\[
X_{1}=-1,\ X_{3}=X_{3}^{\ast}=X_{4}^{-1}\neq1.
\]
The $2\times2$ Jordan block relates to one propagating Bloch mode with
$k=\pi/L$ and zero group velocity, and one non-Bloch linearly diverging
eigenmode. The pair of real eigenvalues $X_{3}$ and $X_{4}=X_{3}^{-1}$ relate
to the pair of evanescent modes at $\omega=\omega_{g}$.

By contrast, at the frequency of photonic band edge $g$ in Fig. \ref{AB_DR}(b)
related to the stack of isotropic layers, there is no evanescent modes. The
canonical Jordan form of the respective transfer matrix coincides with
$\ \tilde{T}_{22}$ in (\ref{Jordan}), where%
\[
X_{1}=X_{2}=-1.
\]
Each of the two $2\times2$ identical Jordan blocks relates to one propagating
Bloch mode with $k=\pi/L$ and zero group velocity, as well as one non-Bloch
linearly diverging eigenmode. The two Jordan blocks of $\tilde{T}_{22}$
correspond to two different polarizations of light.

At the frequency $\omega_{a}$ in Fig. \ref{DRd_4}, the canonical Jordan form
of the respective transfer matrix is $\tilde{T}_{21}$, where%
\[
X_{1}=-1,\ X_{3}=X_{4}^{\ast},\ \left\vert X_{3}\right\vert =\left\vert
X_{4}\right\vert =1.
\]
The double eigenvalue $X_{1}=-1$ of the $2\times2$ Jordan block relates to one
propagating Bloch mode with $k=\pi/L$ and zero group velocity, and one
non-Bloch linearly diverging eigenmode. The pair of complex eigenvalues
$X_{3}$ and $X_{4}=X_{3}^{\ast}$ relate to the pair of propagating modes
having equal and opposite group velocities and belonging to the dispersion
curve other than the one containing the point $a$.

The canonical form $\tilde{T}_{3}$ in (\ref{Jordan}) relates to a $k-\omega$
diagram with stationary inflection point. Such a stationary point cannot be
realized in a reciprocal periodic stack at normal light propagation
\cite{PRB03,PRE03}.

Of particular interest here is the case (\ref{w = k^4}) of degenerate band
edge. At the degenerate band edge frequency $\omega_{d}$, the canonical Jordan
form of the respective transfer matrix is $\tilde{T}_{4}$ in (\ref{Jordan})%
\begin{equation}
T_{L}\sim\tilde{T}_{4}=\left[
\begin{array}
[c]{cccc}%
X & 1 & 0 & 0\\
0 & X & 1 & 0\\
0 & 0 & X & 1\\
0 & 0 & 0 & X
\end{array}
\right]  ,\text{ \ where \ }X=\pm1 \label{T_4}%
\end{equation}
More specifically, in the case shown in Fig. \ref{DRd_4}(b)
\[
X=X_{d}=e^{ik_{d}L}=-1,~\text{at }\omega=\omega_{d}.
\]
The matrix (\ref{T_4}) presents a single $4\times4$ Jordan block and has a
single eigenvector, corresponding to the propagating eigenmode with $k=\pi/L$
and zero group velocity. The other three solutions for the Maxwell equation
(\ref{ME4}) at $\omega=\omega_{d}$ are non-Bloch eigenmodes diverging as $z$,
$z^{2}$, and $z^{3}$, respectively.

If the frequency $\omega$ deviates from the stationary point (\ref{SP}), the
transfer matrix $T_{L}$ become diagonalizable with the canonical Jordan form
$\tilde{T}_{1}$ in (\ref{Jordan}). The perturbation theory relating the
non-Bloch eigenmodes at the frequency of degenerate band edge to the Bloch
eigenmodes in the vicinity of this point is presented in \cite{WRM}.

\subsection{Symmetry conditions for the existence of degenerate band edge}

Not any periodic stack can develop a degenerate band edge, defined in
(\ref{w = k^4}). Some fundamental restrictions can be derived from symmetry
considerations. These restrictions stem from the fact that at the frequency
$\omega_{d}$ of degenerate band edge, the transfer matrix $T_{L}$ must have
the Jordan canonical form (\ref{T_4}). Such a matrix cannot be reduced to a
block-diagonal form, let alone diagonalized. Therefore,\medskip

\emph{- a necessary condition for the existence of degenerate band edge is
that the symmetry of the periodic array does not impose the reducibility of
the transfer matrix }$T_{L}$\emph{ to a block-diagonal form.}\medskip

The above condition does not imply that the transfer matrix $T_{L}$ must not
be reducible to a block-diagonal form at \emph{any} frequency $\omega$ on the
$k-\omega$ diagram. Indeed, at a general frequency $\omega$, the matrix
$T_{L}$ is certainly reducible, and even diagonalizable. The strength of
\emph{the symmetry imposed} reducibility is that it leaves no room for
exceptions, such as the frequency $\omega_{d}$ of degenerate band edge, where
the transfer matrix $T_{L}$ must not be reducible to a block-diagonal form.
Therefore, in the case of symmetry imposed reducibility, the very existence of
degenerate band edge is ruled out. Observe that in most periodic layered
structures, the symmetry of the periodic array \emph{does require} the matrix
$T_{L}$ to be similar to a block-diagonal matrix at all frequencies. In all
these cases, the stack symmetry is incompatible with the existence of
degenerate band edge on the $k-\omega$ diagram.

Let us apply the above criterion to some specific cases.

In periodic stacks of isotropic layers, the Maxwell equations for the waves
with the $x$- and the $y$ - polarizations are identical and decoupled,
implying that the respective transfer matrix can be reduced to the
block-diagonal form%
\begin{equation}
\tilde{T}_{L}=\left[
\begin{array}
[c]{cccc}%
T_{11} & T_{12} & 0 & 0\\
T_{21} & T_{22} & 0 & 0\\
0 & 0 & T_{11} & T_{12}\\
0 & 0 & T_{21} & T_{22}%
\end{array}
\right]  . \label{TL(2=2)}%
\end{equation}
The two identical blocks in (\ref{TL(2=2)}) correspond to two different
polarizations of light. The characteristic polynomial $F_{4}(X)$ of the
block-diagonal matrix (\ref{TL(2=2)}) factorizes into a product of two
identical second degree polynomials related to electromagnetic waves with the
$x$- and the $y$ - polarizations, respectively
\begin{equation}
F_{4}(X)=F_{2}(X)F_{2}(X). \label{T=T^2}%
\end{equation}
The block-diagonal structure of the matrix (\ref{TL(2=2)}) rules out the
existence of degenerate band edge in periodic stacks of isotropic layers. In
fact, the transfer matrix $T_{L}$ in this case can only have the following two
canonical forms%
\[
\tilde{T}_{1}=\left[
\begin{array}
[c]{cccc}%
X & 0 & 0 & 0\\
0 & X^{-1} & 0 & 0\\
0 & 0 & X & 0\\
0 & 0 & 0 & X^{-1}%
\end{array}
\right]  ,~~\tilde{T}_{22}=\left[
\begin{array}
[c]{cccc}%
\pm1 & 1 & 0 & 0\\
0 & \pm1 & 0 & 0\\
0 & 0 & \pm1 & 1\\
0 & 0 & 0 & \pm1
\end{array}
\right]  ,
\]
compatible with (\ref{TL(2=2)}). The case $\tilde{T}_{1}$ relates to a general
frequency, while the case $\tilde{T}_{22}$ relates to a photonic band edge,
like the one shown in Fig. \ref{AB_DR}(b).

Let us now turn to the situation where all or some of the layers of the
periodic stack are birefringent. The in-plane dielectric anisotropy
(\ref{eps trnsv}) may allow for degenerate band edge on the $k-\omega$
diagram, but not automatically.

Let us start with the simplest periodic array in which all anisotropic layers
of the stack have aligned in-plane anisotropy. The term "aligned" means that
one can choose the directions of the in-plane Cartesian axes $x$ and $y$ so
that the permittivity tensors in all layers are diagonalized simultaneously.
In this setting, the Maxwell equations for the waves with the $x$- and the $y$
- polarizations are still separated, implying that the respective transfer
matrix can be reduced to the block-diagonal form%
\begin{equation}
\tilde{T}_{L}=\left[
\begin{array}
[c]{cccc}%
T_{11} & T_{12} & 0 & 0\\
T_{21} & T_{22} & 0 & 0\\
0 & 0 & T_{33} & T_{34}\\
0 & 0 & T_{43} & T_{44}%
\end{array}
\right]  . \label{TL(2x2)}%
\end{equation}
The two blocks in (\ref{TL(2x2)}) correspond to the $x$ and $y$ polarization
of light. The forth degree characteristic polynomial of the block-diagonal
matrix (\ref{TL(2x2)}) factorizes into the product%
\begin{equation}
F_{4}(X)=F_{x}(X)F_{y}(X), \label{T=TxTy}%
\end{equation}
where $F_{x}(X)$ and $F_{y}(X)$ are independent second degree polynomials
related to electromagnetic waves with the $x$- and the $y$ - polarizations,
respectively. The typical $k-\omega$ diagram in this case will be similar to
that shown in Fig. \ref{DRd_4}(a) with two separate curves related to two
linear polarizations of light. Again, the block-diagonal structure of the
matrix (\ref{TL(2x2)}) rules out the existence of degenerate band edge in
periodic stacks with aligned anisotropic layers. The transfer matrix $T_{L}$
in this case can only have the following two canonical forms%
\[
\tilde{T}_{1}=\left[
\begin{array}
[c]{cccc}%
X_{1} & 0 & 0 & 0\\
0 & X_{1}^{-1} & 0 & 0\\
0 & 0 & X_{2} & 0\\
0 & 0 & 0 & X_{2}^{-1}%
\end{array}
\right]  ,~~\tilde{T}_{22}=\left[
\begin{array}
[c]{cccc}%
\pm1 & 1 & 0 & 0\\
0 & \pm1 & 0 & 0\\
0 & 0 & X & 0\\
0 & 0 & 0 & X^{-1}%
\end{array}
\right]  ,
\]
compatible with (\ref{TL(2x2)}). The case $\tilde{T}_{1}$ relates to a general
frequency, while the case $\tilde{T}_{22}$ relates to a photonic band edge.

As we have seen, the presence of anisotropic layers may not necessarily lift
the symmetry prohibition for the degenerate band edge, because the symmetry of
the periodic array may still be incompatible with the canonical Jordan form
(\ref{T_4}). Generally if the space symmetry group $G$ of the layered
structure includes a mirror plane $m_{||}$ parallel to the $z$ direction, this
would guarantee the reducibility of the matrix $T_{L}$ to a block-diagonal
form. Indeed, a standard line of reasoning gives that if $m_{||}\in G$, and
the $y$ axis is chosen perpendicular to the mirror plane $m_{||}$, then the
waves with the $x$- and the $y$ - polarizations have different parity with
respect to the symmetry operation of reflection and, therefore, are decoupled.
The latter leads to reducibility of the matrix $T_{L}$ to the block-diagonal
form (\ref{TL(2x2)}). Thus, a formal necessary condition for the existence of
degenerate band edge on the $k-\omega$ diagram can be written as follows%
\begin{equation}
m_{||}\notin G. \label{DBE crit}%
\end{equation}

None of the common periodic layered structures satisfies this criterion and,
therefore, none of them can develop the degenerate band edge. For example,
even if anisotropic layers are present, but the anisotropy axes in all
anisotropic layers are either aligned, or perpendicular to each other, the
symmetry group $G$ of the stack still has the mirror plane $m_{||}$, which
guarantees the separation of the $x$- and the $y$ - polarizations and the
reducibility of the respective transfer matrix $T_{L}$ to the block-diagonal
form (\ref{TL(2x2)}). The only way to satisfy the condition (\ref{DBE crit})
and, thereby, to allow for degenerate band edge on the $k-\omega$ diagram, is
to have at least two misaligned anisotropic layers in a unit cell with the
misalignment angle being different from $0$ and $\pi/2$, as shown in the
example in Fig. \ref{ABA}.

Observe that the presence of $B$ layers in the periodic array in Fig.
\ref{ABA} is also essential, unless the two anisotropic layers $A_{1}$ and
$A_{2}$ have different thicknesses or are made of different anisotropic
materials. In Fig. \ref{ABA}, the layers $A_{1}$ and $A_{2}$ differ only by
their orientation in the $x-y$ plane, but otherwise, they are identical. In
such a case, if the $B$ layers are removed, the point symmetry group of the
periodic stack in Fig. \ref{ABA} rises from $D_{2}$ to $D_{2h}$ acquiring the
glide mirror plane $m_{||}$. This, according to the criterion (\ref{DBE crit}%
), imposes the reducibility of the matrix $T_{L}$ to the block-diagonal form
(\ref{TL(2x2)}), regardless of the misalignment angle between the adjacent $A$
layers. The symmetry imposed reducibility rules out the possibility of the
degenerate band edge (\ref{w = k^4}). The $k-\omega$ diagram of the periodic
stack in Fig. \ref{ABA} with the $B$ layers removed is shown in Fig.
\ref{DRd_4}(d).

In the numerical example considered in the next section, the $B$ layers are
simply empty gaps of certain thickness $D_{B}$ between the adjacent double
layers $A_{1}-A_{2}$. The misalignment angle is chosen $\pi/4$. By changing
the thickness $D_{B}$ of the gap $B$, one can change the $k-\omega$ diagram of
the periodic stack, as shown in Fig. \ref{DRd_4}. Similar effect can be
achieved by changing the misalignment angle between the adjacent $A$ layers.

\section{Transmission resonance in the vicinity of degenerate band edge}

\subsection{Scattering problem for periodic semi-infinite stack}

To solve the scattering problem for a plane monochromatic wave incident on a
finite stack of anisotropic layers we use the following standard approach
based on the $4\times4$ transfer matrix.

Let $\Psi_{I}\left(  z\right)  $, $\Psi_{R}\left(  z\right)  $, and $\Psi
_{P}\left(  z\right)  $, be the incident, reflected, and passed plane waves in
vacuum. Allowing for general elliptic polarization of the waves, we have%
\[
\Psi_{I}\left(  z\right)  =\left[
\begin{array}
[c]{c}%
A_{x}\\
A_{y}\\
-A_{y}\\
A_{x}%
\end{array}
\right]  e^{i\frac{\omega}{c}z},\ \Psi_{R}\left(  z\right)  =\left[
\begin{array}
[c]{c}%
R_{x}\\
R_{y}\\
R_{y}\\
-R_{x}%
\end{array}
\right]  e^{-i\frac{\omega}{c}z},\ \Psi_{P}\left(  z\right)  =\left[
\begin{array}
[c]{c}%
P_{x}\\
P_{y}\\
-P_{y}\\
P_{x}%
\end{array}
\right]  e^{i\frac{\omega}{c}z}.
\]
The respective domains of definition are%
\begin{equation}%
\begin{array}
[c]{c}%
z\leq0\text{, \ for \ }\Psi_{I}\left(  z\right)  \text{\ and }\Psi_{R}\left(
z\right)  ,\\
D\leq z\text{, \ for \ }\Psi_{P}\left(  z\right)  .
\end{array}
\label{Psi DD}%
\end{equation}
where $D$ is the stack thickness. The field inside the stack is denoted by
$\Psi_{T}\left(  z\right)  $. Except for the stationary points (\ref{SP}) on
the $k-\omega$ diagram, $\Psi_{T}\left(  z\right)  $ can be decomposed into a
superposition of the four Bloch solutions (\ref{Bloch}) of the Maxwell
equation (\ref{ME4})%
\begin{equation}
\Psi_{T}\left(  z\right)  =\sum_{k}\Psi_{k}\left(  z\right)  ,\ \ 0\leq z\leq
D. \label{Psi T}%
\end{equation}
This representation is meaningful if the periodic stack contains a significant
number of unit cells $L$. Otherwise, if there are just a few layers in the
stack, the representation (\ref{Psi T}) is formally valid, but not
particularly useful.

The boundary conditions at the two slab/vacuum interfaces are%
\begin{equation}
\Psi\left(  0\right)  =\Psi_{I}\left(  0\right)  +\Psi_{R}\left(  0\right)
,\;\Psi\left(  D\right)  =\Psi_{P}\left(  D\right)  . \label{BC2}%
\end{equation}
The transfer matrix $T_{S}$ of a periodic stack is%
\begin{equation}
T_{S}=\left(  T_{L}\right)  ^{N}. \label{T_N}%
\end{equation}
The relation%
\begin{equation}
\Psi\left(  D\right)  =T_{S}\Psi\left(  0\right)  , \label{Psi D=T Psi 0}%
\end{equation}
together with the pair of boundary conditions (\ref{BC2}) allow to express
both the reflected wave $\Psi_{R}$ and the wave $\Psi_{P}$ passed through the
slab, in terms of a given incident wave $\Psi_{I}$ and the elements of the
transfer matrix $T_{S}$. This also gives the transmittance/reflectance
coefficients of the slab defined as%
\begin{equation}
\tau_{D}=\frac{S_{P}}{S_{I}}=\frac{\left\vert \Psi_{P}\left(  D\right)
\right\vert ^{2}}{\left\vert \Psi_{I}\left(  0\right)  \right\vert ^{2}%
},\;\rho_{D}=\frac{S_{R}}{S_{I}}=\frac{\left\vert \Psi_{R}\left(  0\right)
\right\vert ^{2}}{\left\vert \Psi_{I}\left(  0\right)  \right\vert ^{2}}.
\label{tN,rN}%
\end{equation}
where $S=cW$ is the Poynting vector of the respective wave. In the case of a
lossless stack%
\[
\;\;\tau_{D}+\rho_{D}=1.
\]

The field distribution $\Psi_{T}\left(  z\right)  $ inside the slab is found
using either of the following expressions%
\begin{equation}
\Psi_{T}\left(  z\right)  =T\left(  z,0\right)  \left[  \Psi_{I}\left(
0\right)  +\Psi_{R}\left(  0\right)  \right]  =T\left(  z,D\right)  \Psi
_{P}\left(  D\right)  ,\ \ 0\leq z\leq D. \label{PsiT(z)}%
\end{equation}

The above procedure is commonly used for the frequency-domain analysis of
periodic and non-periodic layered structures involving anisotropic and/or
gyrotropic layers (see, for example, \cite{Tmatrix},\cite{Abdul00}%
,\cite{Abdul99},\cite{Manda03},\cite{PRE01,PRB03,PRE03,PRE05} and references therein).

In addition to the field distribution inside the slab, we are also interested
in its eigenmode composition. The latter is particularly important since it
allows to explain the fundamental difference between Fabry-Perot resonance in
the vicinity of a degenerate band edge and a similar resonance in the vicinity
of a regular band edge. Throughout this section we consider only the first
transmission resonance, closest to the respective band edge.

\subsection{Field composition at the frequency of transmission resonance}

\begin{figure}[tbph]
\scalebox{0.75}{\includegraphics[viewport=0 0 500 250,clip]{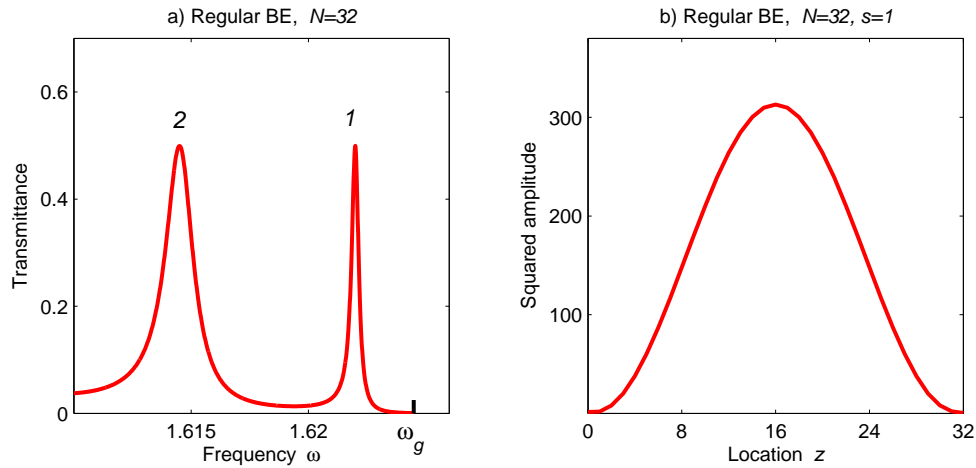}}
\caption{Transmission resonance in the vicinity of regular photonic band edge
(BE) $g$ in Fig. 6(a) in the stack composed of $32$ unit cells $L$. \ a)
Transmission dispersion at the frequency range including the regular BE $g$
and two closest transmission resonances. b) Smoothed field intensity
distribution (59) inside the stack at the frequency $\omega_{1}$ of the first
transmission resonance $1$.}
\label{tf32gy_Am32gy}
\end{figure}

\begin{figure}[tbph]
\scalebox{0.75}{\includegraphics[viewport=0 0 500 250,clip]{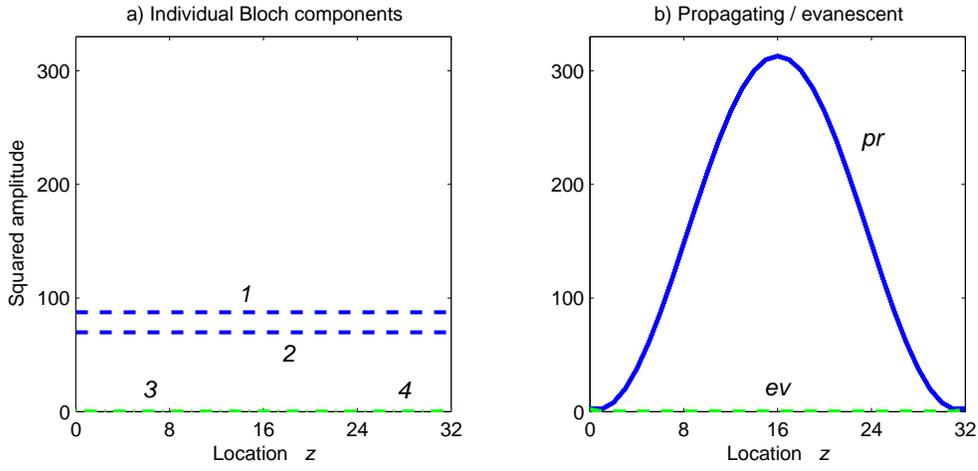}}
\caption{The Bloch composition of the resonant field in Fig. 9(b)
corresponding to the transmission peak $1$ in Fig. 9(a). a) Square moduli of
each of the four Bloch components; $1$ and $2$ are forward and backward
propagating components. b) Square moduli of the combined contribution of two
propagating components ($pr$), and the combined contribution of two evanescent
components ($ev$). The evanescent modes contribution is negligible.}
\label{Amg32y_2}
\end{figure}

In the case of transmission resonance in a periodic stack of isotropic layers,
the resonance field inside the stack is a simple standing wave composed of two
propagating Bloch modes with opposite group velocity (see Eqs. (\ref{k_s})
through ((\ref{W_g}) and the comments therein). The introduction of anisotropy
in itself does not change qualitatively the resonance picture, as shown in
Fig. \ref{tf32gy_Am32gy}. The only difference is that the electromagnetic
field $\Psi_{T}\left(  z\right)  $ inside the stack can now have both
propagating and evanescent components. But at the frequency of a transmission
resonance, the contribution of the evanescent components is negligible, as
shown in Fig. \ref{Amg32y_2}. So, basically, one can still see the resonance
field inside the stack as a simple standing wave composed of a pair of
propagating modes with greatly enhanced amplitude, compared to that of the
incident wave. The formulas (\ref{k_s}) through ((\ref{W_g}) still apply here,
provided that the number $N$ of unit cells in the stack is not too small.

In Fig. \ref{Amg32y_2}(a) we show the squared amplitudes (intensity
distributions)%
\begin{equation}
\left\vert \Psi_{j}\left(  z\right)  \right\vert ^{2},\ j=1,2,3,4,
\label{W 1,2,3,4}%
\end{equation}
of the individual Bloch components of the resulting field $\Psi_{T}\left(
z\right)  $ at the frequency of the first transmission resonance near the
regular band edge $g$ in Fig. \ref{DRd_4}(a). The numbers 1 and 2 designate
the forward and backward propagating components of the standing wave. The
evanescent contributions 3 and 4 are negligible. In Fig. \ref{Amg32y_2}(b) we
show the squared amplitudes of the combined contribution of the pair of
propagating waves ($pr$), and the combined contribution of the pair of
evanescent waves ($ev$)%
\begin{equation}
\left\vert \Psi_{pr}\left(  z\right)  \right\vert ^{2}=\left\vert \Psi
_{1}\left(  z\right)  +\Psi_{2}\left(  z\right)  \right\vert ^{2},\ \left\vert
\Psi_{ev}\left(  z\right)  \right\vert ^{2}=\left\vert \Psi_{3}\left(
z\right)  +\Psi_{4}\left(  z\right)  \right\vert ^{2}. \label{Wpr, Wev}%
\end{equation}
Obviously,%
\begin{equation}
\Psi_{T}\left(  z\right)  \approx\Psi_{pr}\left(  z\right)  .
\label{Psi_T, Psi_pr}%
\end{equation}

Now let us turn to the case of transmission resonance in the vicinity of
degenerate band edge $d$ in Fig. \ref{DRd_4}(b). There are several features
that sharply distinguish this case from the similar transmission resonance
near the regular band edge $g$ in Fig. \ref{DRd_4}(a).

First of all, for a given number $N$ of unit cells in the stack ($N=32$), the
field intensity in Fig. \ref{Amdny}(b) is by several orders of magnitude
higher, compared to that in Fig. \ref{tf32gy_Am32gy}(b), in spite of the close
similarity of all numerical parameters of the respective periodic structures
(see the Appendix). In the case of transmission resonance in the vicinity of
degenerate band edge (DBE), the resonance field intensity increases as $N^{4}%
$, while in the case of a regular band edge (BE), the field intensity is
proportional to $N^{2}$.

The second distinction is that in the case of degenerate band edge (DBE), the
field intensity near the slab boundaries at $z=0$ and $z=D$, increases as%
\begin{equation}
\text{DBE case: }\left\vert \Psi_{T}\left(  z\right)  \right\vert ^{2}\propto
z^{4},\ \left(  D-z\right)  ^{4}. \label{Psi DBE}%
\end{equation}
By contrast, in the case of a regular band edge (BE), the field intensity near
the stack boundaries rises at a much slower rate%
\begin{equation}
\text{Regular BE case: }\left\vert \Psi_{T}\left(  z\right)  \right\vert
^{2}\propto z^{2},\ \left(  D-z\right)  ^{2}, \label{Psi BE}%
\end{equation}
which is characteristic of a regular standing wave composed of two propagating components.

\begin{figure}[tbph]
\scalebox{0.75}{\includegraphics[viewport=0 0 500 250,clip]{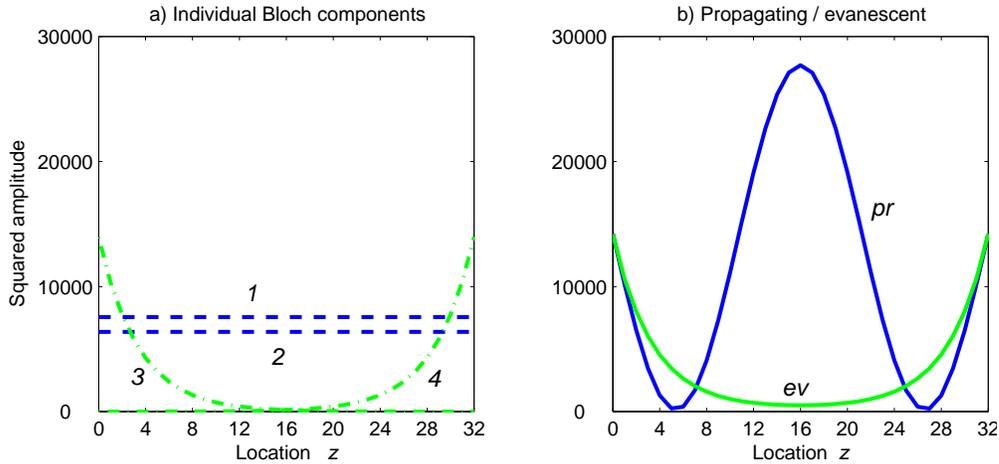}}
\caption{The Bloch composition of the resonant field in Fig. 8(b)
corresponding to the transmission peak $1$ in Fig. 7. a) Square moduli of each
of the four Bloch components, $1$ and $2$ are forward and backward propagating
components. b) Square moduli of the combined contribution of two propagating
components ($pr$), and the combined contribution of two evanescent components
($ev$). Propagating and evanescent contributions interfere destructively at
the slab boundaries.}
\label{Amd32y_2}
\end{figure}

Finally, in the case of a regular band edge, the transmission resonance field
$\Psi_{T}\left(  z\right)  $ is a standing wave composed of two propagating
Bloch modes with opposite group velocities. The evanescent modes do not
participate in the formation of the resonance field, as clearly seen in Fig.
\ref{Amg32y_2}. By contrast, in the case of transmission resonance in the
vicinity of DBE, the role of evanescent components in the formation of the
resonance field is absolutely crucial. Indeed, as shown in Fig. \ref{Amd32y_2}%
, amplitude of the propagating and evanescent components are comparable in
magnitude. More importantly, the combined contribution $\Psi_{pr}\left(
z\right)  $ of the two propagating components does not even resemble a
standing wave with the nodes at the slab boundary, as was the case in Fig.
\ref{Amg32y_2}(b). Instead, comparing Fig. \ref{Amdny}(b) and Fig.
\ref{Amd32y_2}(b) we see that at the slab boundaries at $z=0$ and $z=D$. the
propagating and evanescent components interfere destructively, almost
canceling each other%
\[
\text{at }z=0\text{ and }z=D\text{: }\Psi_{pr}\left(  z\right)  \approx
-\Psi_{ev}\left(  z\right)  ,
\]
while the individual Bloch components (\ref{W 1,2,3,4}) remain huge. We remind
that in all cases, the intensity of the incident wave $\Psi_{I}$\ is unity.%

Qualitatively, such a bizarre resonance behavior in the vicinity of the
degenerate band edge can be characterized as follows. As we already mentioned,
at the frequency $\omega_{d}$ of degenerate band edge, three of the four
solutions for the Maxwell equation (\ref{ME4}) are non-Bloch eigenmodes
diverging as $z$, $z^{2}$, $z^{3}$. Although the frequency $\omega_{1}$ of the
transmission cavity resonance is slightly different from $\omega_{d}$ and,
therefore, all four eigenmodes are Bloch waves, the close proximity of
$\omega_{1}$ to $\omega_{d}$ causes all the abnormalities seen in Fig.
\ref{Amg32y_2}. To describe these behavior in mathematically consistent way,
one should start with the one solutions at $\omega=\omega_{d}$ as zero
approximation. Then using the perturbation theory for the non-diagonalizable
transfer matrix $T_{L}\left(  \omega_{d}\right)  $, one can derive an
asymptotic theory for the case of large \thinspace$N$. The perturbation theory
for the degenerate band edge $d$ was developed in \cite{WRM}. It turns out
that near the slab boundaries at $z\ll D$ and $\left(  D-z\right)  \ll D$, the
field $\Psi_{T}\left(  z\right)  $ at the transmission resonance frequency
$\omega_{1}$ is well approximated by quadratically diverging non-Bloch
eigenmode, corresponding to $\omega=\omega_{d}$. It qualitatively explains an
extremely rapid growth (\ref{Psi DBE}) of the resonance field inside the slab
as one moves away from either slab boundary. It also explains the sharp
dependence (\ref{W_d}) of the resonance field intensity on the number $N$ of
unit cells in the stack.\bigskip

\textbf{Acknowledgment and Disclaimer:} Effort of A. Figotin is sponsored by
the Air Force Office of Scientific Research, Air Force Materials Command,
USAF, under grant number FA9550-04-1-0359.

\section{Appendix}

\subsection{Numerical parameters of layered arrays}

The periodic array in Fig. \ref{ABA}. has three layers in a unit cell $L$, of
which two ($A_{1}$ and $A_{2}$) are anisotropic and have the same thickness
$D_{A}$. The third layer $B$ is isotropic with the thickness $D_{B}$. In our
numerical simulations, the $B$ layers are empty gaps of variable thickness.

The anisotropic layers $A_{1}$ and $A_{2}$ are made of the same lossless
dielectric material and have the same thickness. The respective dielectric
permittivity tensor is%
\[
\hat{\varepsilon}=\left[
\begin{array}
[c]{ccc}%
\varepsilon_{A}+\delta\cos2\varphi & \delta\sin2\varphi & 0\\
\delta\sin2\varphi & \varepsilon_{A}-\delta\cos2\varphi & 0\\
0 & 0 & \varepsilon_{zz}%
\end{array}
\right]  ,\;
\]
where $\delta$ describes the magnitude of in-plane anisotropy, while the angle
$\varphi$ defines the orientation of the anisotropy axes of the respective
layer in the $x-y$ plane. The most critical parameter of the periodic
structure in Fig. \ref{ABA} is the misalignment angle%
\begin{equation}
\varphi=\varphi_{1}-\varphi_{2} \label{phi}%
\end{equation}
between the adjacent anisotropic layers $A_{1}$ and $A_{2}$. This angle
determines the symmetry of the periodic array and, eventually, what kind of
$k-\omega$ diagram it can display. It is important for our purposes that the
misalignment angle is different from $0$ and $\pi/2$. In our numerical
simulations we set%
\[
\varphi_{1}=0,\varphi_{2}=\pi/4,
\]
and use the following expressions for the dielectric permittivity of the $A$
layers%
\begin{equation}
\hat{\varepsilon}_{A1}=\left[
\begin{array}
[c]{ccc}%
\varepsilon_{A}+\delta & 0 & 0\\
0 & \varepsilon_{A}-\delta & 0\\
0 & 0 & \varepsilon_{zz}%
\end{array}
\right]  ,\;\hat{\varepsilon}_{A2}=\left[
\begin{array}
[c]{ccc}%
\varepsilon_{A} & \delta & 0\\
\delta & \varepsilon_{A} & 0\\
0 & 0 & \varepsilon_{zz}%
\end{array}
\right]  . \label{eps ABA}%
\end{equation}
In our numerical simulations we set%
\[
\varepsilon_{A}=13.\,\allowbreak61,\ \delta=12.\,\allowbreak4.
\]
The numerical value of $\varepsilon_{zz}$ is irrelevant.

In the case of the periodic array of isotropic layers in Fig. \ref{AB_DR}(a),
we set%
\[
\varepsilon_{A}=3.78,\ \delta=0,\ \ D_{A}=D_{B}=0.5\times L.
\]

\subsection{Description of plots}

In all plots of the field intensity distribution we, in fact, plotted the
following physical quantity%
\begin{equation}
\left\langle \left\vert \Psi\left(  z\right)  \right\vert ^{2}\right\rangle
=\left\langle \vec{E}\left(  z\right)  \cdot\vec{E}^{\ast}\left(  z\right)
+\vec{H}\left(  z\right)  \cdot\vec{H}^{\ast}\left(  z\right)  \right\rangle
_{L}, \label{Sm Int}%
\end{equation}
which is the squared field amplitude averaged over local unit cell. The real
electromagnetic energy density $W\left(  z\right)  $ is similar to $\left\vert
\Psi\left(  z\right)  \right\vert ^{2}$. Both of them are strongly oscillating
functions of the coordinate $z$ with the period of oscillations coinciding
with the unit cell length $L$. Thus, the quantity (\ref{Sm Int}) can be
interpreted as the smoothed field intensity distribution, with the correction
coefficient of the order of unity.

In all plots, the wave number $k$ and the frequency $\omega$ are expressed in
units of $L^{-1}$ and $cL^{-1}$, respectively.


\begin{thebibliography}{99}                                                                                               %


\bibitem {Yariv}A. Yariv and Pochi Yeh. \textsl{Optical Waves in Crystals}.
(\textquotedblright A Wiley-Interscience publication\textquotedblright, 1984).

\bibitem {Yeh}Pochi Yeh. \textquotedblright\textsl{Optical Waves in Layered
Media}\textquotedblright, (Wiley, New York, 1988).

\bibitem {Chew}Weng Cho Chew. \textquotedblright\textsl{Waves and Fields in
Inhomogeneous Media}\textquotedblright, (Van Nostrand Reinhold, New York, 1990).

\bibitem {CR  Meloni 03}A. Melloni, F. Morichetti, M. Maritelli.
\textsl{Linear and nonlinear pulse propagation in coupled resonator slow-wave
optical structures. }Optical and Quantum Electronics \textbf{35}, 365 (2003)

\bibitem {Manda03}A. Mandatori, C. Sibilia, M. Centini, G. D'Aguanno, M.
Bertolotti, M. Scalora, M. Bloemer, and C. M. Bowden. \textsl{Birefringence in
one-dimensional finite photonic band gap structure}. J. Opt. Soc. Am. B
\textbf{20}, 504 (2003).

\bibitem {SL Scal1}M. Scalora, R. J. Flynn, S. B. Reinhardt, R. L. Fork, M. J.
Bloemer, M. D. Tocci, C. M. Bowden, H. S. Ledbetter, J. M. Bendickson, J. P.
Dowling, R. P. Leavitt. \textsl{Ultrashort pulse propagation at the photonic
band edge: Large tunable group delay with minimal distortion and loss}. Phys.
Rev. E54, \#2, R1078 (1996).

\bibitem {SL Scal2}M. Bloemer, K. Myneni, M. Centini, M. Scalora, and G.
D'Aguanno. \textsl{Transit time of optical pulses propagating through a finite
length medium}. Phys. Rev. E\textbf{65}, 056615 (2002).

\bibitem {SL Joann}M. Soljacic, S. Johnson, S. Fan, M. Ibanescu, E. Ippen, and
J. D. Joannopoulos. \textsl{Photonic-crystal slow-light enhancement of
nonlinear phase sensitivity}. J. Opt. Soc. Am. B., \textbf{19}, \#9, 2052 (2002).

\bibitem {CR Yariv04}J. Poon, J. Scheuer, Y. Xu, and A. Yariv.
\textsl{Designing coupled-resonator optical waveguide delay lines}. J. Opt.
Soc. Am. B, Vol. \textbf{21}, No. 9 (2004).

\bibitem {Tmatrix}D. W. Berreman. J. Opt. Soc. Am. A\textbf{62}, 502--10 (1972).

\bibitem {Abdul00}I. Abdulhalim. \textsl{Analytic propagation matrix method
for anisotropic magneto-optic layered media}, J.Opt. A: Pure Appl.
Opt.\textbf{2}, 557 (2000).

\bibitem {Abdul99}I. Abdulhalim. \textsl{Analytic propagation matrix method
for linear optics of arbitrary biaxial layered media}, J.Opt. A: Pure Appl.
Opt. \textbf{1}, 646 (1999).

\bibitem {PRE01}A. Figotin, and I. Vitebsky. \textsl{Nonreciprocal magnetic
photonic crystals}. Phys. Rev. \textbf{E}63, 066609 (2001).

\bibitem {PRB03}A. Figotin, and I. Vitebskiy. \textsl{Electromagnetic
unidirectionality in magnetic photonic crystals}. Phys. Rev. \textbf{B}67,
165210 (2003).

\bibitem {PRE03}A. Figotin, and I. Vitebskiy. \textsl{Oblique frozen modes in
layered media}. Phys. Rev. \textbf{E}68, 036609 (2003).

\bibitem {PRE05}J. Ballato, A. Ballato, A. Figotin, and I. Vitebskiy.
\textsl{Frozen light in periodic stacks of anisotropic layers}. Phys. Rev.
\textbf{E}71, (2005).

\bibitem {WRM}A. Figotin, and I. Vitebskiy. \textsl{Slow light in photonic
crystals }(Topical Review). Submitted to "Waves in Random and Complex Media"
(arXiv:physics/0504112 v2 19 Apr 2005).

\bibitem {LODE}E. Coddington and R. Carlson. \textsl{Linear Ordinary
Differential Equations}. (SIAM, Philadelphia, 1997).

\bibitem {Linalg}R. Bellman. \textsl{Introduction to Matrix Analysis}. (SIAM.
Philadelphia, 1997)

\bibitem {LanTi}Lancaster P. and Tismenetsky M., \textsl{The Theory of
Matrices}, Academic Press, 1985.

\bibitem {Wilk}Wilkinson, J., \textsl{The Algebraic Eigenvalue Problem},
Oxford University Press, 1996.
\end{thebibliography}
\end{document}